\documentclass[article,authoryear,mst]{beg_32}            

\usepackage[hang]{footmisc}
\usepackage{comment}

\fancypagestyle{plain}{%
  \fancyhf{}
  \fancyhead[R]{\small {\it \jname}, x(x):\thepage--\pageref{LastPage} (\myyear\today)}
  \fancyfoot[R]{\small\bf\thepage }
  \fancyfoot[L]{\fottitle}
  }
\renewcommand{\dmyy}{23}
\renewcommand{\myyear}{2023}
\renewcommand{\today}{}
\begin{document}

\volume{Volume x, Issue x, \myyear\today}
\title{Characterization of a capillary driven flow in microgravity by means of optical technique}
\titlehead{Characterization of a capillary driven flow in microgravity by means of optical technique}
\authorhead{D. Fiorini, L. Carbonnelle, A. Simonini, J. Steelant, D. Seveno and M. A. Mendez}
\corrauthor[1,2]{Domenico Fiorini}
\author[1]{Louis Carbonnelle}
\author[1]{Alessia Simonini}
\author[3]{Johan Steelant}
\author[2]{David Seveno}
\author[1]{Miguel Alfonso Mendez}
\corremail{domenico.fiorini@vki.ac.be}
\corraddress{von Karman Institute for Fluid Dynamics, Waterloosesteenweg 72, Sint-Genesius-Rode, 1640, Belgium}
\address[1]{von Karman Institute for Fluid Dynamics, Waterloosesteenweg 72, Sint-Genesius-Rode, 1640, Belgium}
\address[2]{KU Leuven, Dept. of Materials Engineering, Leuven 3001, Belgium}
\address[3]{ESTEC-ESA, Keplerlaan 1, Noordwijk, The Netherlands}

\dataO{02/20/2023}
\dataF{mm/dd/yyyy}

\abstract{The motion of a gas-liquid interface along a solid wall is influenced by the capillary forces resulting from the interface's shape and its interaction with the solid, where it forms a dynamic contact angle. Capillary models play a significant role in the management of cryogenic propellants in space, where surface tension dominates the behaviour of gas-liquid interfaces. Yet, most empirical models have been derived in configurations dominated by viscous forces. In this study, we experimentally investigate the wetting of a low-viscosity, highly wetting fluid in a reduced gravity environment. Our setup consisted of a transparent and diverging U-tube in which capillary forces sustain the liquid motion. Combining Particle Image Velocimetry (PIV) and high-speed backlighting visualization, the experimental campaign allowed for measuring the interface evolution and the velocity field within the liquid under varying gravity levels. This work reports on the preliminary results from the image velocimetry and shows that the velocity profile within the tube is close to parabolic until a short distance from the interface. Nevertheless, classic 1D models for capillary rise face difficulties reproducing the interface dynamics, suggesting that the treatment of the surface tension in these problems must be reviewed.}

\keywords{Dynamic Wetting in Microgravity, Time-Resolved Particle Image Velocimetry, Interface dynamics in capillary tubes}

\maketitle

\section{Introduction}
Modelling the dynamics of a gas-liquid interface along a solid surface is fundamental in wetting and dewetting (e.g. inkjet printing), coating flows (e.g. slot die coating), and capillary-driven flows (e.g. capillary tubes). These applications have spurred much of the research on the topic and have often been analyzed using semi-empirical or theoretical correlations that link the contact angle at the wall to the contact-line dynamics.

Classic empirical laws relate the contact angle to the velocity of the contact line through the Capillary number (e.g. \cite{voinovhydrodynamics,jiangcorrelation,bracke1989kinetics,kistlerhydrodynamics}), while other authors proposed more complex correlations, in the form of differential equations, to take into account the time history of the contact-line dynamics (e.g. \cite{ting_perlin,bianliquid}).

Dynamic contact angle correlations are essential to predict interface dynamics. However, \citet{fiorinieffect} recently showed that classic quasi-stationary models of the interface shape fail in situations where inertia plays a significant role. This is the case in many applications in microgravity, such as the sloshing of space propellants. A critical case study for cryogenic propellant management in microgravity is the axial sloshing of a highly wetting liquid in axisymmetrical containers. This test case was analyzed by \citet{utsumilow} and \citet{luppes2009numerical}. Their results demonstrated that the modelling of surface tension strongly impacts the prediction of liquid displacement. Similarly, \citet{lidynamic,lidynamics} conducted a CFD analysis of the oscillations of the gas-liquid interface in a cylindrical tank upon a step reduction in gravity and compared the results with the experimental findings of \citet{lidynamics}. Their results showed that the dynamic contact angle models significantly impact the rising velocity of the contact line. 

While these studies highlight the relevance of capillary forces on the motion of the liquid interface in the absence of gravity and complex geometries, other studies have addressed the fundamentals of dynamic wetting in microgravity using dedicated capillary rise experiments. \citet{stangecapillary} performed a capillary rise experiment for a perfect wetting fluid on a large cylindrical tube in a drop tower to reproduce microgravity conditions. They showed that the interface reorientation and capillary rise consist of three phases. The first is dominated by the inertia of the meniscus acceleration. Viscous effects dominate the second, while capillary forces dominate the third phase. The extension of these three phases depends on the duration of the experiments. \citet{wanginfluence} conducted a parametric study of capillary rise in the same configuration to examine the impact of the contact angle and tube size. The two parameters proved to be coupled, with the interface in large tubes producing faster capillary rise at an earlier stage for low contact angles. This behaviour was interpreted as the result of a competition between viscous and inertial effects, with the former dominating the capillary rise in small-diameter tubes and the latter affecting the initial stage of the rise in large (inertia-dominated) tubes.

The aforementioned studies provide essential data for validating and deriving dynamic contact angle correlations through 1D models of the capillary rise, such as the one developed by \cite{stangecapillary}. However, these experiments provide limited information on capillary-dominated conditions as they neglect the role of the flow field beneath the interface, which, in turn, impacts the dynamic contact angle. The velocity field underneath an advancing meniscus has been observed by a few experiments in small capillary tubes, where the motion of the liquid can be assumed to be viscous-dominated. For example, \citet{nasarek2008flow} and \citet{waghmare2012experimental} performed a micro-PIV experiment in a vertical channel and analyzed the advancing motion of a meniscus of ethanol and ethylene glycol, respectively. Both works showed that the common assumption of Poiseuille flow holds until a certain distance from the interface, below which the velocity profile flattens, and radial components appear. Similarly, \citet{ratzanalysis} analyzed the flow of a quasi-capillary water channel in inertia-dominated conditions and showed that rolling motion under the interface could be produced in advancing and receding conditions. 

In this work, we use parabolic flights to study experimentally the capillary rise and the interface dynamics of highly wetting fluids in microgravity. The experimental conditions achieved through the flights provide a larger duration of the capillary-dominated phase compared to previous studies and allow for characterizing the relationship between the interface shape, its motion, and the underlying flow field at different gravity levels.

The analyzed configuration consists of a diverging U-tube in a saturated gas environment. This prevents evaporation, which can have a significant impact on the wetting process, as shown in  \cite{waghmare2012experimental}. We use particle image velocimetry to measure the velocity field beneath the meniscus and high-speed visualization in back-lighting, combined with image processing, to measure the interface evolution. In this article, we report on the PIV measurements of a representative test case. Moreover, we present a simple 1D modelling of the interface dynamics on the U tube and compare its prediction with the available experimental data.

The results highlight the role of the capillary forces near the interface and their impact on the capillary rise of highly wetting liquids. These results are particularly interesting to the management of space propellants, which are also characterized by low (static) contact angles.

\section{Experimental Facility}\label{sec:experimentalfacility}

\begin{figure}[h]
       \centering
        \includegraphics[height=6.5cm,clip]{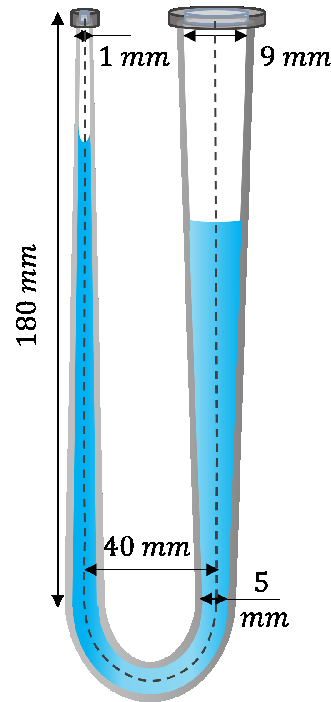}
    \includegraphics[height=6.cm, trim={1cm 1cm 1cm 1cm},  clip]{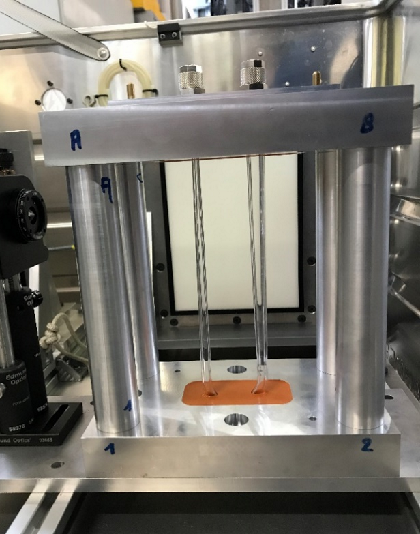}
 \caption{The Figure on the left shows a schematic of the DUT test cell with the characteristic dimensions. The Figure on the right shows a picture of the DUT mounted on a rigid structure to avoid resonance vibrations in the quartz.}
 \label{fig:DUT}
\end{figure}

The diverging U-tube (DUT) configuration analyzed in this work is shown in Fig. \ref{fig:DUT}. The experiments were carried out during the 78th ESA parabolic flight campaign. 

In the DUT, the liquid reservoir used in classic capillary rise experiments is replaced by the larger side of the U-tube. This minimizes the impact of liquid sloshing occurring during a parabolic flight. Moreover, capillary forces act on both sides of the liquid column, creating a capillary pressure drop that pulls the interface in opposite directions. This smooths the acceleration profile compared to the step response in experiments with a single driving interface, reducing the inertia-dominated phase and increasing the impact of capillary forces on the interface motion. Finally, the lack of a tube entrance reduces viscous losses due to sudden tube constriction, minimizing the effect of viscous forces and reducing the duration of the viscous-dominated phase identified by \citet{stangecapillary}. 

Figure \ref{fig:DUT} shows the installation of the DUT quartz tube on its robust structure in aluminium. The structure consists of two plates joined by four cylindrical beams via M8 screws. A resin is poured into two pockets of the structure to clamp the tube. The system is designed to allow optical access while having sufficient rigidity to avoid vibration, which could result in the cracking of the quartz. The complete structure with the U tube and the resin weighs five kg.

\begin{figure}[h]
    \centering
    \includegraphics[height=5.3cm,clip]{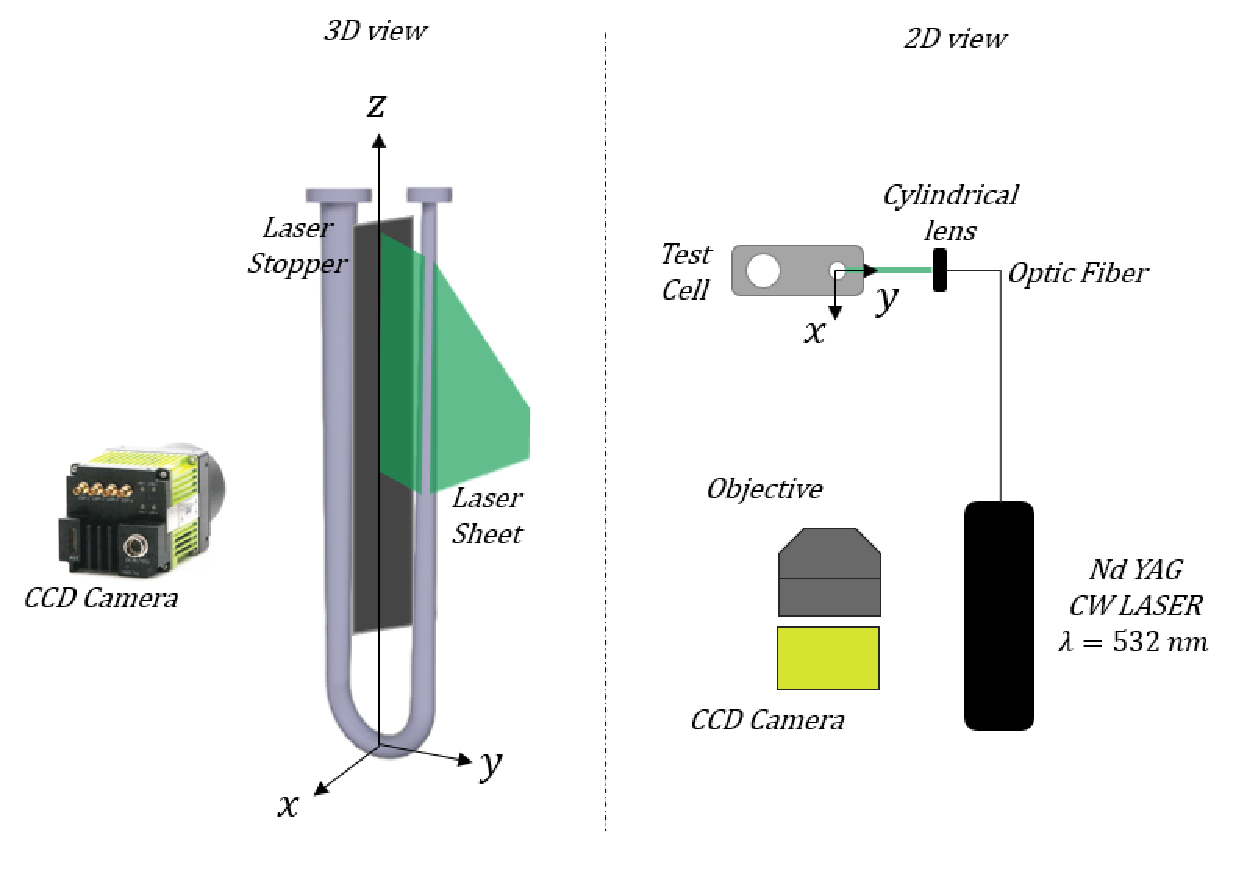}
    \includegraphics[height=5.3cm,clip]{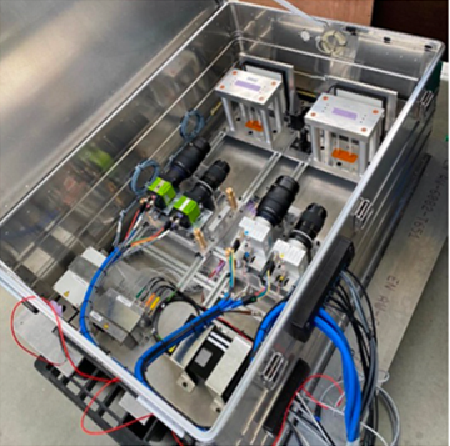}
 \caption{The figures on the left show a schematic of the optical set-up for the PIV measurements. The back-lighting is performed using a LED screen behind the U-tube's aluminium frames. The Figure on the right shows the installation on the parabolic flight breadboard: two experiments were carried out simultaneously.}
 \label{fig:schemeandsetup}
\end{figure}

The two sides of the tube are connected with a flexible junction. This prevents leakages and creates a closed environment in which saturated conditions are reached before the experiments begin. The flexible is connected to the plate with two fast connectors on the top. The assembly is then installed on a larger breadboard containing the entire experimental equipment, as shown in Fig.  \ref{fig:schemeandsetup}. A separate electronic rack contains all the electronic devices necessary for the experiment. These include acquisition cards for the acceleration and pressure signals connected to a laptop, accelerometer conditioner, laser controller and camera controller connected to a desktop. 

Two experiments were performed simultaneously, with a camera acquiring images on each side of the two tubes. Both experiments used HFE7200 ($\text{C}_\text{4} \text{F}_\text{9} \text{O} \text{C}_\text{2} \text{H}_\text{5}$) produced by 3M Novec as test fluid. The fluid physical properties have been characterized by \citet{rausch2015density} and evaluated at the experiment temperature (298K). The gas phase is constituted by air saturated with HFE7200 vapor. The motion of each interface is recorded with a high-speed camera (model JAY SP-12000-CXP4), acquiring grey-scale images at 300 fps. The active region of all cameras is restricted to the central region of size $4096 \times n_{px}$ pixels where the lateral size $n_{px}$ is optimized according to the tube geometry to achieve the highest acquisition frequency allowed by the camera. All cameras mount objectives with 105mm focal lenses and are positioned to acquire the full motion of the interface while spanning the largest possible tube length. The final pixel size corresponds to $0.013\mu m$.

As shown in Fig. \ref{fig:schemeandsetup} on the left, the PIV is carried out using a planar laser sheet produced on one side of the tube. A laser stopper is introduced to avoid optical disturbances on the other side of the tube. The high speed in backlighting conditions is carried out using a LED screen behind the DUTs. However, in this work, we solely focus on the PIV campaign. The fluid was seeded with FMR-1.3 1-5 $\mu m$ Red Fluorescent Miscospheres from Cospheric. The particles were illuminated with a continuous laser source (LMX-532 from Oxxius) with a wavelength of 532 nm and a maximum power of 800 mW. 

The experiment recordings start with the Pilot's announcement that the plane reached 40 degrees inclination. Shortly after the announcement, the gravity levels drop from approximately 1.8$g$ to approximately 0$g$, and the two liquid interfaces begin moving along the axis of the tube due to the unbalanced capillary action at the interfaces. The image processing consists of several stages. First, the images are compensated for the vibrations of the camera objective, which appears as displacement of the image in the vertical (z) and lateral (x) direction (max 4 px.) and without significant image rotation. To compensate for the vibrations, we detect the outer edges of the tube in each image via standard canny edge detection and compute the necessary x and y correction by iteratively minimizing the difference of the global overlap of the image with a reference image acquired before the experiment start. In the second step, the PIV images are pre-processed using the Proper Orthogonal Decomposition (POD)-based background removal introduced by \citet{mendezpod}. The optical distortion due to the tube's curved surface was corrected using the method proposed by \citet{darzi2017optical}. This approach corrects the radial distortion given the distance of the camera from the tube center, the tube radius and the refraction index of the materials involved. The method was validated in \cite{fiorinieffect} in similar size cylindrical tube. The undistorted image is then mapped at the original camera resolution using bicubic image interpolation.

Finally, the PIV interrogation is carried out using OpenPIV processing tool (\cite{liberzonjakirkham}) and windows with a 2:1 aspect ratio, given that velocity component v (along y) is much larger than the velocity component u (along x). We use a Dynamic Region of Interest (DROI) to follow the gas-liquid interface across the available Field of View (FOV). This allows focusing the analysis always in the region below the interface. The DROI is identified by performing standard gradient-based edge detection on the POD-filtered images to detect the element of interest in the image (e.g. the interface). Spurious points and outliers are eliminated through both masking of the tube solid surfaces and unsupervised outlier detection using the Local Outlier Factor (LOF) by \citet{breunig2000lof}. While this approach allows identifying the region of the image where PIV can be applied, the light refraction at the interface of the PIV images prevents extracting the interface contact angle with high accuracy, for which the backlight technique is necessary. The results of the contact angle analysis will be presented in the extended version of this work. The PIV images are processed using adaptive iterative multigrid interrogation (\cite{scaranoadvances}) to retrieve the velocity fields. The initial window size is set to $128\times64$ px and reduced over two steps to $32\times16$ px. Outliers are removed based on the signal-to-noise ratio measured in terms of peak-to-peak in the cross-correlation map.

\section{Modeling of the capillary rise }

\begin{figure}[h]
\centering
\includegraphics[height=3cm,clip]{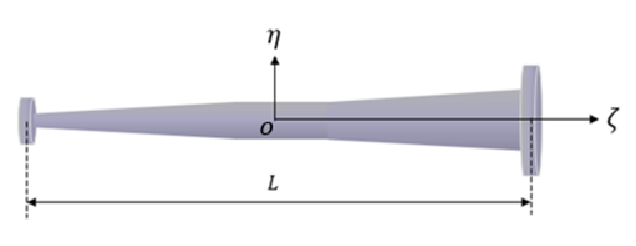}
\caption{Equivalent cylindrical coordinate system for DUT where gentle curvature of the tube is assumed.}
\label{subfig:coordinates2D}
\end{figure}

We describe the motion of the liquid's center of mass inside the DUT with a 1D integral model balancing the acceleration of the center of mass with the friction, gravity and the interface capillary pressure drop. We derive the model assuming a gentle curvature of the tube and a cylindrical coordinate system $(\zeta,\eta)$ with $\zeta$ aligned with the tube axis and $\eta$ aligned along the tube radius as illustrated in Figure \ref{subfig:coordinates2D}.  

We integrate the contributions of all forces along the liquid domain to obtain the following second-order non-linear ODE:
\begin{equation}
    \rho V_l \ddot{Y} = - F_g(Y,a(t))+F_{v}(\dot{Y},Y)+F_{c}(\dot{Y},Y)\,,
    \label{eq:dutmodel}
\end{equation} where $Y$, $\dot{Y}$ and $\ddot{Y}$ are respectively the position, the velocity and the acceleration of the liquid column's center of mass along the tube axis $\zeta \in[-L/2,L/2]$. On the right-hand side, $F_g$ denotes the gravitational force acting to the center of mass of the liquid column, $F_v$ denotes the viscous shear applied to the liquid column, and $F_{c}$ denotes the net result of the capillary pressure due to the two gas-liquid interfaces on the opposite sides of the liquid column. $V_l$ is the volume of liquid within the tube and $a(t)=[a_x(t),a_y(t),g(t)]$ is the local acceleration of the tube.

We denote with $z_A(Y)$ and $z_B(Y)$ the position of the two interfaces along the curvilinear coordinate $\zeta$. The gravity term in \eqref{eq:dutmodel} results from the contribution of both the height difference $|z_A(Y)|-|z_B(Y)|$ and the local acceleration $a(t)$. 

The viscous force term in \eqref{eq:dutmodel} is computed from two contributions: the first is the integration of the viscous term in the Navier Stokes equations, here denoted as $\tau$, along the liquid column
$z_A(Y)-z_B(Y)$ and the radial direction $\eta$; the second is an empirical correction to account for the viscous losses at the tube's bending. These two terms reads as follows:

\begin{equation}
\label{vis}
    F_{v}(\dot{Y},Y)=\int_{z_A(Y)}^{z_B(Y)}\int_{0}^{R(\zeta)}  \tau\,  2\pi \eta\, d\eta \,d\zeta+C_U\,.
\end{equation} 
where value of $C_U$ was computed from the correction proposed by \cite{ito1969laminar}:
\begin{equation}
   C_U=8\pi\mu V_0\left(\pi D_U\right)\left(0.1033 De^{0.5}\left(\left(1+\frac{1.729}{De}\right)^{0.5}-\left(\frac{1.729}{De}\right)^{0.5}\right)^{-3}\right)
    \,,
\end{equation} 
with $De$ the Dean number computed as $De=Re(V_0)\sqrt{{2R}/{D_U}}$, with $V_0=V(\zeta=0,t)$ the velocity at the centerline, $Re(V_0)$ the local Reynolds number based on the centerline velocity and $D_U$ the distance between the tube axes. 

Concerning the first term in \eqref{vis}, assuming axisymmetric flow, the diffusion term $\tau$ can be written as 
\begin{equation}
    \tau = \mu\left[\frac{1}{\eta}\frac{\partial v}{\partial\eta}+\frac{\partial^2 v}{\partial \eta^2}+\frac{\partial^2 v}{\partial \zeta^2}\right] 
\end{equation} where $v = v(\zeta,\eta, t)$ is the instantaneous axial velocity at the coordinate $\zeta,\eta$. In normal gravity conditions ($g=9.8 m/s^2$) and for small diameters, classic capillary rise models assume a Poiseuille velocity profile everywhere in the channel (see for example \cite{zhonganalytic}). 
With this simplification, tested in this work for the DUT configuration, the velocity profile can be written as
\begin{equation}
   v= v_p(\zeta,\eta, V(\zeta,t))=2V(\zeta,t)\left(1-\frac{\eta^2}{R(\zeta)^2}\right)\,.
    \label{eq:poiseuille}
\end{equation} where $V$ denotes the mean velocity and $R(\zeta)$ the radius of the channel as a function of the channel axial coordinate. 

The capillary term is modelled as the net sum of the pressure drop on the two interfaces, which in the analyzed wetting configuration pulls the liquid towards opposite sides. The pressure drop results from the integration of Laplace equation for each interface and reads 

\begin{equation}
F_{c}(\dot{Y},Y)  = \int_{0}^{R(z_A(Y))}\Delta P_A(\eta) 2\pi \eta d\eta-\int_{0}^{R(z_B(Y))}\Delta P_B(\eta) 2\pi \eta d\eta\,.\label{eq:capillaryforcegeneral}
\end{equation}

In the absence of mass transfer and Marangoni effects, the capillary pressure drops $\Delta P(\eta,t)_{A|B}$ become

\begin{equation}
    \Delta P(\eta,t) = \sigma\nabla\cdot \mathbf{n}\,
    \label{eq:shape}
\end{equation} where $\nabla\cdot \mathbf{n}$ is the interface curvature, with $\nabla$ the nabla operator and $\mathbf{n}$ the local normal vector. If the interface is treated as spherical, the curvature becomes $\nabla\cdot \mathbf{n}=2/R_s$, with $R_s$ the sphere's radius. With this assumption, introducing the dependency on the apparent and dynamic contact angle $\theta$, one has $R_s(\zeta)\approx R(\zeta)/cos(\theta)$ and thus the capillary pressure drops on each side reads:

\begin{equation}
    \Delta P_{A} =  \frac{2\sigma}{R(z_A(Y))} cos(\theta_A) \quad \mbox{and}\quad \Delta P_{B} =   \frac{2\sigma}{R(z_B(Y))} cos(\theta_B)
\end{equation}

\section{Results and Discussion}

\begin{figure}[h!]
    \centering
    \includegraphics[height = 4.7cm, trim={0cm 3cm 0cm 3cm},  clip]{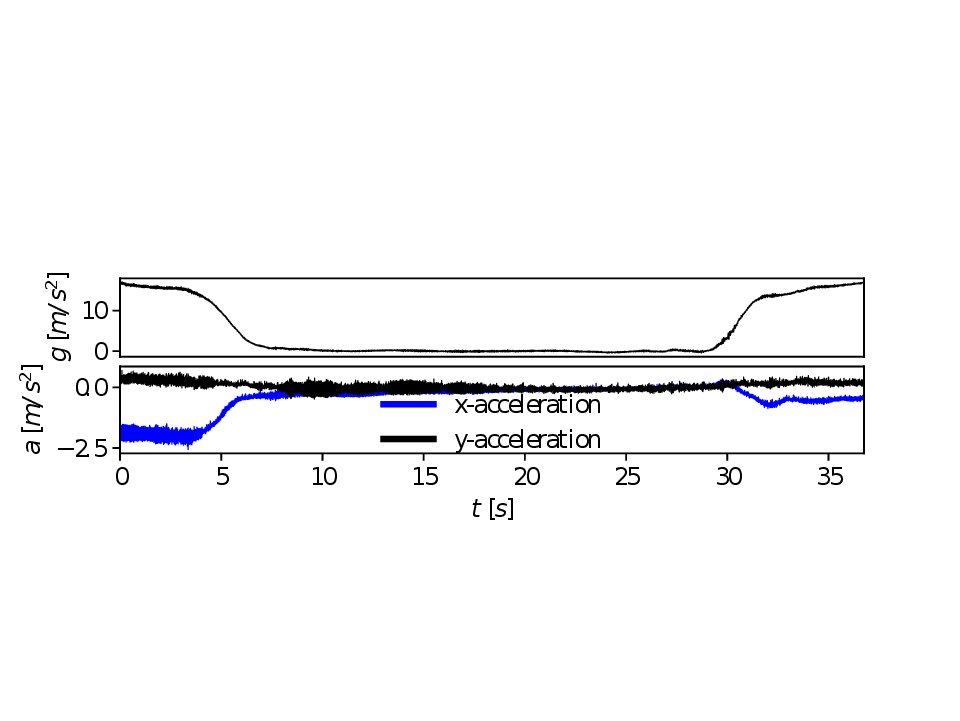}
    \caption{Acceleration profile for the selected experiment}
    \label{fig:acceleration profile}
\end{figure}

Fig. \ref{fig:acceleration profile} shows the acceleration profiles for the presented experiments during the $\approx 38$s constituting a parabola. The figure on the top shows the envelope of (vertical) gravitational acceleration while the figure on the bottom report on the lateral accelerations (see Fig. \ref{fig:schemeandsetup}).

\begin{figure}[h!]
    \centering
    \includegraphics[height = 14cm, trim={0cm 0cm 4cm 0cm},clip,  angle=-90]{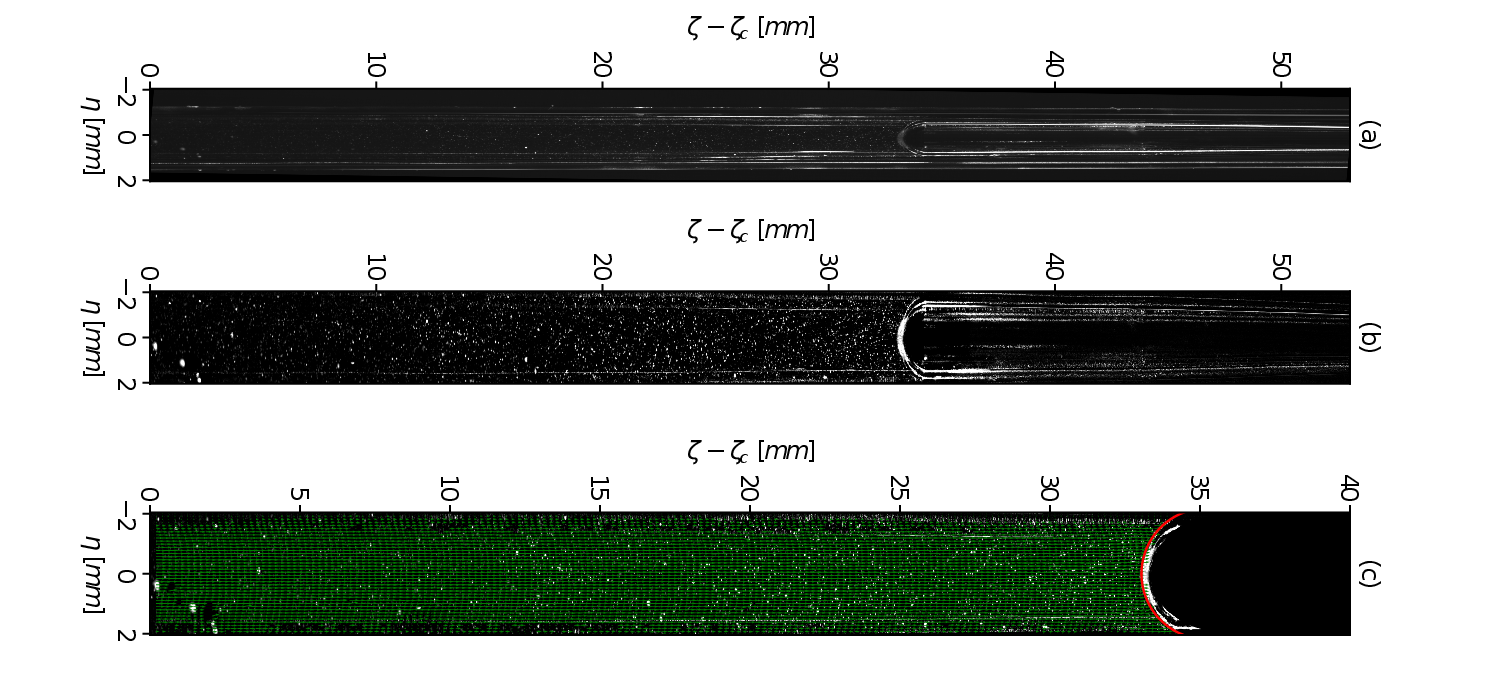}
    \caption{ From top to bottom, the Figure shows the raw PIV images and the result from the background removal. All figures are rotated for plotting purposes.}
    \label{fig:introPIV}
\end{figure}

\begin{figure}[h]
\centering
    \includegraphics[height=6cm,]{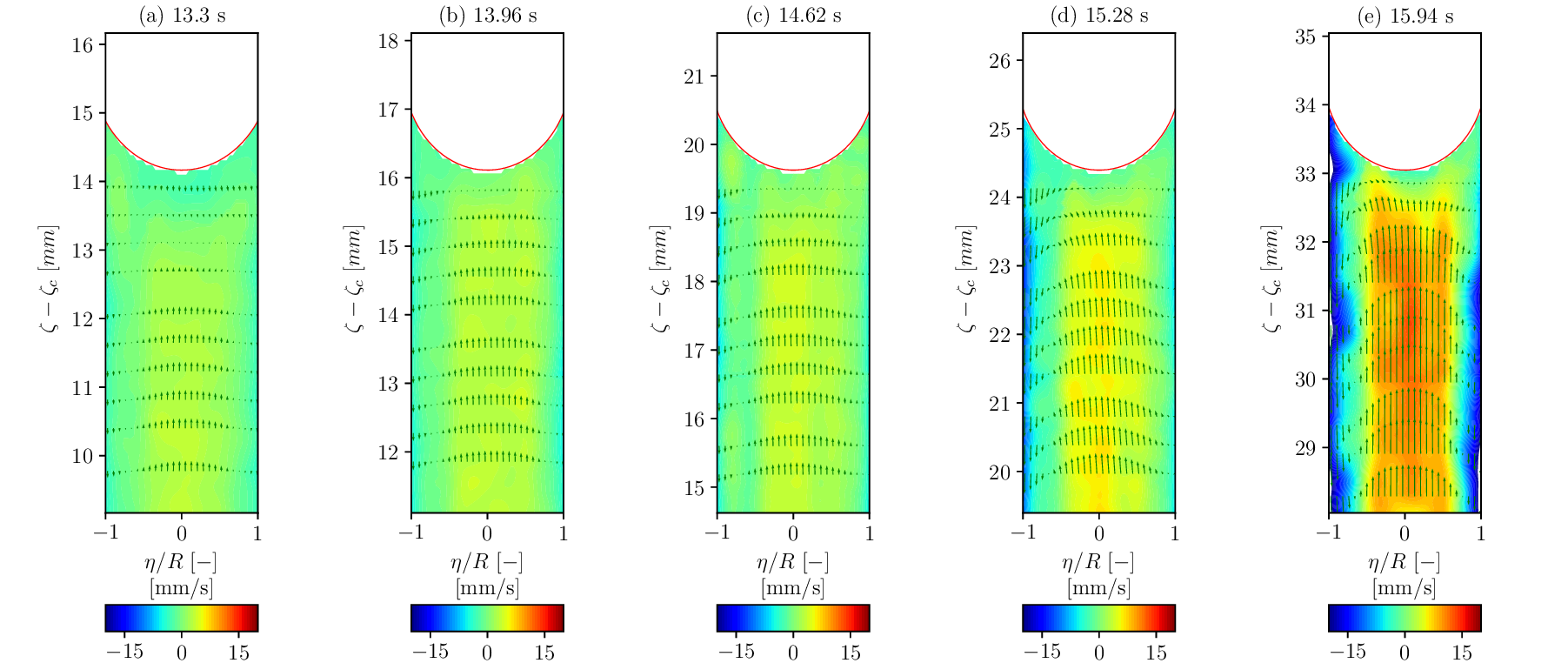}
\caption{From left to right, the plot shows the axial velocity profile in a moving reference frame with rising velocity $\dot{Y}$ and at extracted locations in the axis of the tube. Each plot also shows the contour of the velocity magnitude beneath the advancing meniscus.}
\label{fig:velocityfield_selected}
\end{figure}

The micro-gravity phase begins approximately at $t=10$ s and ends at $t=30$s. This is preceded and followed by a hypergravity phase. The lateral accelerations are non-negligible during the hypergravity phases and can have an important effect also during the microgravity phase.  
Figure \ref{fig:introPIV}a shows a sample image acquired during the experiment. All images are pre-processed using the POD background removal; the result for the snapshot in \ref{fig:introPIV}a is shown in \ref{fig:introPIV}b. The velocity field is computed using the multigrid approach with final-size windows (32, 16). The result is shown in Fig \ref{fig:velocityfield_selected} for an extracted axial location near the interface. 

In what follows, we focus on the velocity field close to the interface  up to 5 mm below the interface. The measured velocity fields in this area are shown in Fig. \ref{fig:velocityfield_selected} using a moving reference frame with rising velocity $\dot{Y}$. The red curve in this plot shows the result of the interface detection described in the previous section.
These results show that the liquid accelerates towards the contact line as it approaches the interface. This lateral acceleration is due to mass conservation: since the interface shape remains unvaried and hence all its points move at a constant velocity, the flow must adapt from a nearly parabolic profile to a flat one. A similar result was reported by \citet{waghmare2012experimental} for purely capillary-dominated flow with no evaporation. The classic counter-rotating vortices shown in similar configurations by \citet{davis1974motion} and \citet{ratzanalysis} are not visible in this case. 

At the scale of interest, namely at about the half-width of the channel, the flow appears to be different from the classic wedge flow postulated by \cite{voinovhydrodynamics} in the derivation of its classic theoretical model for the dynamic contact angle. This result might thus question the validity of such a model for the investigated test case.

Concerning the stream-wise variation of the velocity profile, at a distance of approximately 3 mm, the velocity profile appears nearly parabolic in all snapshots. At larger acceleration (see snapshot e) the flow inertia appears to push the validity of the parabolic profile closer to the interface (up to nearly 1 mm from it).

\begin{figure}[h]
    \centering
       \includegraphics[height=6.5cm]{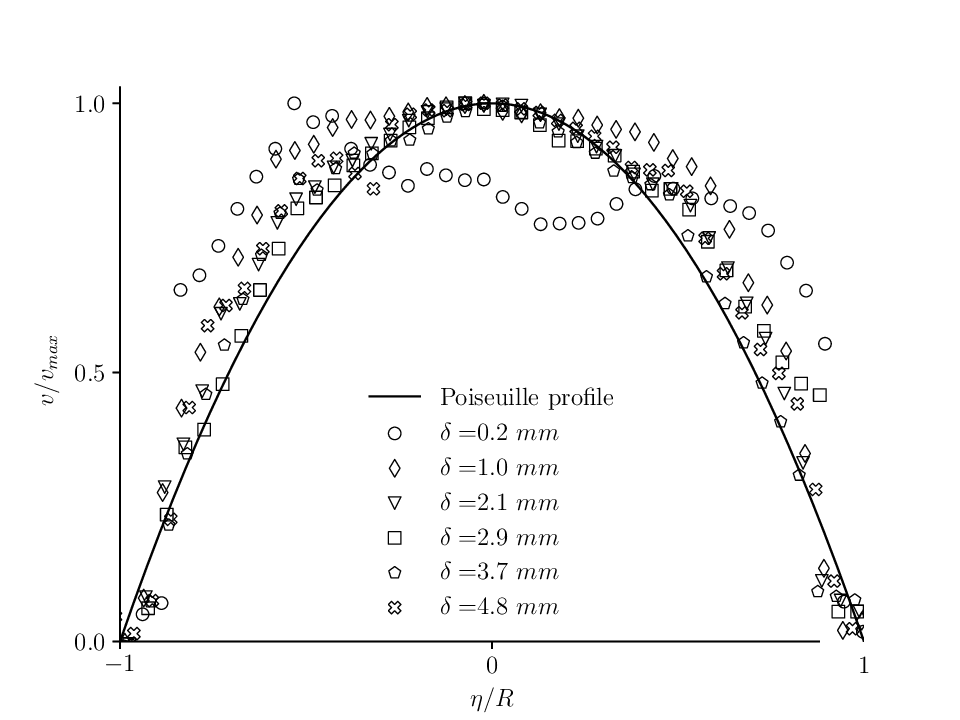}
    \caption{Comparison of experimental radial velocity profiles, sampled at various axial positions along the channel, with theoretical Poiseuille profile.}
    \label{fig:selfsimilar}
\end{figure}

\begin{figure}[h]
\centering
    \includegraphics[height=5cm]{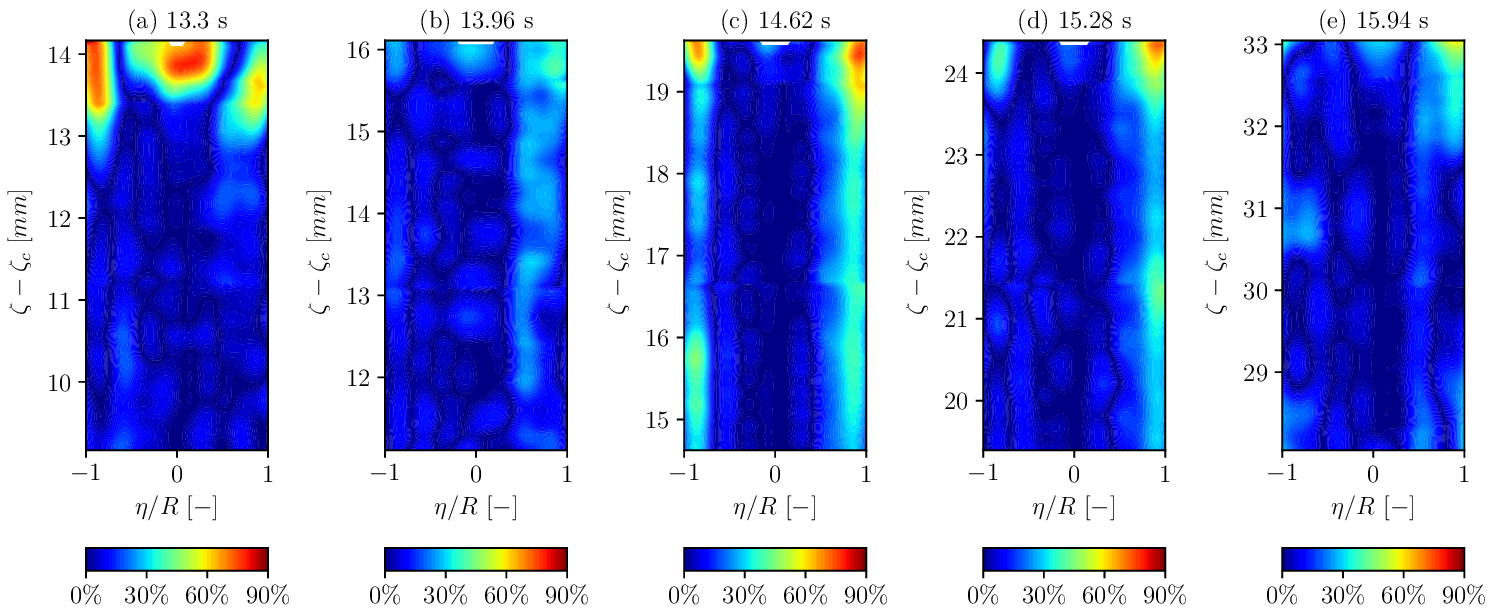}
\caption{From left to right, the contour plots show the difference of the velocity profile with e corresponding Poiseuille profile at each axial location of the channel. The line plots show the increase of disagreement with the velocity profile moving from the bulk flow toward the meniscus.}
\label{fig:parabolicity}
\end{figure}

Figure \ref{fig:selfsimilar} compares the velocity profiles at various distances from the interface in this latter snapshot, characterized by larger inertia. The validity of the self-similar assumption is illustrated by comparing the extracted velocity profiles with the classic parabolic profile from the assumption of Poiseuille flow. The progressive flattening of the velocity profile starts appearing at approximately 1 mm from the interface and the parabolic shape of the velocity profile increasingly deteriorates at closer distances from the interface. Fig. \ref{fig:parabolicity} use contour plots 
 to provide an indication of a lack of similarity with the Poiseuille flow, measured as 

\begin{equation}
    P_{\%}=\frac{|v_{P}-v|}{{v}(0,t)}\times 100\,,
\end{equation} where $v(\eta,\zeta)$ is the measured velocity and $v_{P}(\zeta,\eta,V(\zeta,t))$ is the local velocity under the assumption of parabolic profile in \eqref{eq:poiseuille}. This parameter is defined up to the lowest point of the interface, above which the wetting meniscus begins. The figure shows that at the latest stages of the capillary rise, the velocity field is close to that of a Poiseuille flow until a close distance from the interface.

\begin{figure}[h]
    \centering
       \includegraphics[height=6cm]{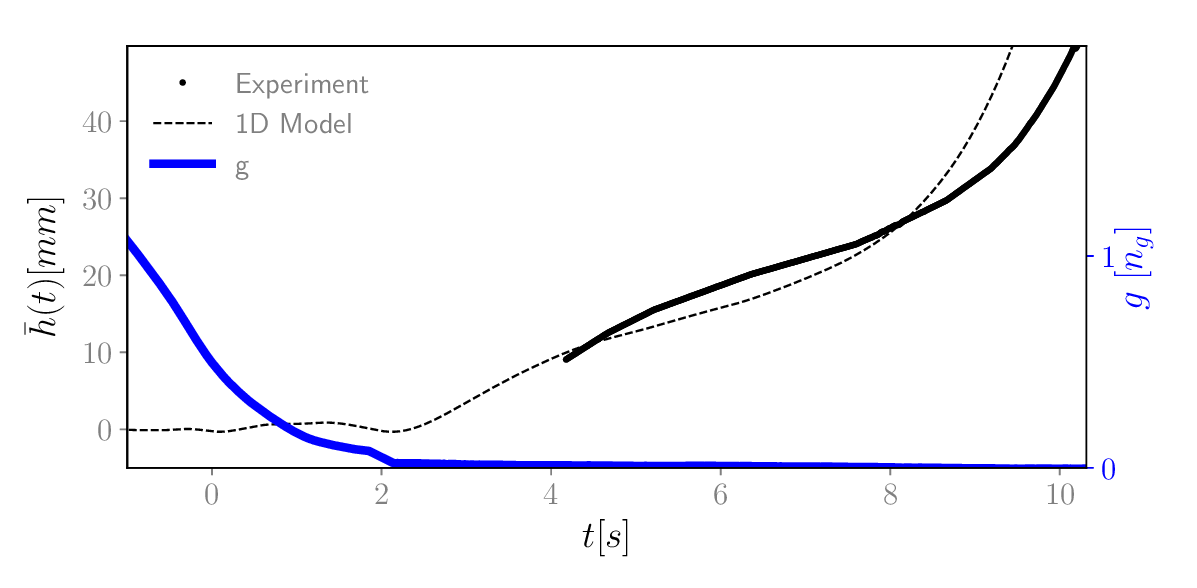}
    \caption{The Figure shows the comparison of the experimental interface position with the prediction achieved by solving \eqref{eq:dutmodel}. The gravity profile corresponding to the time frame examined by the experiment is also shown in the back.}
    \label{fig:model}
\end{figure}
To conclude the analysis of the velocity field, it appears that the assumption of fully developed Poiseuille flow holds until a short distance from the interface, especially when the flow is strongly dominated by inertia as in the last snapshots of Figure \ref{fig:parabolicity}. These results validate the assumption of a self-similar profile in the theoretical model presented in \eqref{eq:dutmodel}, particularly in the computation of the viscous term. Figure \ref{fig:model} compares the model prediction and the experimental data. The axis on the left reports the interface position while the one on the right reports the instantaneous vertical acceleration.

The figure illustrates on the right axis the examined gravity profile and on the left the comparison of the model solution with experimental data on the interface position during the experiment. The model correctly shows the oscillation of the interface velocity due to minor oscillation in the gravitational level at about $t\in[6-8]$s. Since all terms in equation \ref{eq:dutmodel} have a comparable order of magnitude during the interface evolution, the accurate modelling of all terms is relevant to achieve a satisfactory prediction of the interface history. Considering that the assumption of self-similarity appears reasonable, the sources of the discrepancy are found either in modelling the pressure losses due to the bend or the treatment of the capillary forces. The first appears less plausible, given that the employed correlation has been well-calibrated in ground experiments. Therefore, the mismatch is addressed here to the modelling of the capillary term and the impact of the different gravity levels on the interface shape.



\section{Conclusion}

In this work, we investigated the capillary rise dynamics in a divergent U-tube setup under micro-gravity conditions using time-resolved Particle Image Velocimetry (PIV). The configuration produces a flow driven by the difference in capillary forces between the two sides of the tube. We followed both the interface position and the velocity field during the capillary rise and compared the results with a simple 1D integral model based on self-similarity. The objective was to verify the validity of the self-similar velocity profile assumption and test the accuracy of a simplified 1D formulation of the capillary-driven flow.

The PIV measurements showed the assumption of a parabolic velocity profile until a distance of $\approx1$ mm from the interface, thus validating one of the main simplifying assumptions in these flows. The flow profile near the meniscus change from parabolic to nearly flat, with a strong acceleration near the contact line. This suggests that the flow field significantly differs from the wedge flow postulated in the derivation of theoretical models of dynamic wetting, which leads to contact angle correlations of the form $\theta_D=f(Ca)$.


\acknowledgements
D. Fiorini is supported by Fonds Wetenschappelijk Onderzoek (FWO), Project number 1S96120N. This work was supported by the ESA and BELSPO via the Prodex program and by the ESA-GSTP "Physical and Numerical Modeling of cryogenic sloshing for Space applications" (Contract No. 4000129315/19/NL/MG). We would like to thank also Novaspace for their support and the fruitful discussions. Finally, we acknowledge Michel Bavier for the support in the experiment preparation and execution. 











\bibliographystyle{Bibliography_Style}

\bibliography{main}
\end{document}